\documentclass{article} 
\usepackage{iclr2018_conference2,times}
\usepackage{url}

\iclrfinalcopy
\usepackage{latexsym}
\usepackage{amsmath}
\usepackage{amssymb}
\usepackage{hyperref}
\usepackage{setspace}
\usepackage{epsfig}
\usepackage{epstopdf }
\usepackage{graphicx}
\usepackage{times}
\usepackage{color, soul}

\begin{document}

\author{Ricky Fok \thanks{ Corresponding Author}, Aijun An \\
Department of Electrical Engineering and Computer Science\\
York University\\
4700 Keele Street,  Toronto, M3J 1P3, Canada \\
\texttt{\{ricky.fok3@gmail.com\} \{ann@cse.yorku.ca\}} \\
\And
 Xiaogang Wang \\
Department of Mathematics and Statistics \\
York University\\
4700 Keele Street,  Toronto, M3J 1P3, Canada \\
\texttt{stevenw@mathstat.yorku.ca} \\
}

\title{Spontaneous Symmetry Breaking in Deep Neural Networks}

\date{Received: date / Accepted: date}

\maketitle

\begin{abstract}
We propose a framework to understand the unprecedented performance and robustness
of deep neural networks using field theory. Correlations between the
weights within the same layer can be described by symmetries in that layer, and
networks generalize better if such symmetries are broken to reduce the redundancies
of the weights. Using a two parameter field theory, we find that the network
can break such symmetries itself towards the end of training in a process commonly
known in physics as spontaneous symmetry breaking. This corresponds to
a network generalizing itself without any user input layers to break the symmetry,
but by communication with adjacent layers. In the layer decoupling limit applicable
to residual networks \citep{resnet}, we show that the remnant symmetries
that survive the non-linear layers are spontaneously broken. The Lagrangian for
the non-linear and weight layers together has striking similarities with the one in
quantum field theory of a scalar. Using results from quantum field theory we show
that our framework is able to explain many experimentally observed phenomena,, such as training on random labels with zero error \citep{understand}, the information bottleneck, the phase transition out of it and gradient variance explosion \citep{opening}, shattered gradients \citep{shattered}, and many more.

\end{abstract}

\section{Introduction}
Deep Neural Networks have been used in image recognition tasks with great success. The first of its kind, AlexNet \citep{Hinton}, led to many other neural architectures have been proposed to achieve start-of-the-art results in image processing at the time. Some of the notable architectures include, VGG \citep{verydeep}, Inception \citep{goingdeep} and Residual networks (ResNet)\citep{resnet}.

Understanding the inner working of deep neural networks remains a difficult task until now. It has been discovered that the training progress ceases when it goes through an information bottleneck \citep{opening} until learning rate is decreased to a suitable amount then the network under goes a phase transition. Deep networks appear to be able to regularize themselves and able to train on randomly labeled data \citet{understand} with zero training error. The gradients in deep neural networks behaves as white noise over the layers \citet{shattered}. And many other unexplained phenomona. We have found a framework that can explain all the aforementioned results, among many others.

A recent work \citep{warped} showed that the ensemble behavior and binomial path lengths \citep{ensemble} of ResNets can be explained by just a Taylor series expansion to first order in the decoupling limit. They found that the series approximation generates a symmetry breaking layer that reduces the redundancy of weights, leading to a better generalization. Because the ResNet does not contain such symmetry breaking layers in the architecture. They suggest that ResNets are able to break the symmetry by the communication between the layers. Another recent work also employed the Taylor expansion to investigate ResNets \citep{rescon}.


We show that a remnant symmetry subgroup can survive a ReLU layer and the weights can be thought as scalar fields covariant to this remnant symmetry. The emergence of a scalar field is intriguing, perhaps troubling. A scalar field can admit negative Hessian eigenvalues at the loss minimum. Historically, this paradox is solved by promoting a scalar field theory to a quantum field theory incorporating Einstein's special relativity \citep{sred}. But how does this apply to neural networks? We now illustrate that a residual network behaves just like a quantum field theory. In the layer decoupling limit, the norm of the change in feature map representation through each layer has an upper bound, by definition of the layer decoupling. This is similar to relativity where the speed of any particle is bounded from above by the speed of light. In Residual networks, it has been shown that the network can be decomposed into a sum of subnetworks consisting of all possible paths between two layers \cite{ensemble}. In quantum mechanics, the propagation of particle between two points is the sum of all paths connecting those two points. So a residual network can be described by a quantum field theory and the use of a scalar field is justified. In fact, the presence of negative Hessian eigenvalues drives phase transitions in deep neural networks, and such eigenvalues also have been observed \citep{eigen}.

The organization of this paper is as follows. The background on deep neural networks and field theory is given in Section \ref{background}. Section \ref{symNN} shows that remnant symmetries can exist in a neural network and that the weights can be approximated by a scalar field. Experimental results that confirm our theory is given in Section \ref{pred}. A review of field theory is given in Appendix \ref{review}. An explicit example of spontaneous symmetry breaking is shown in Appendix \ref{SSB}.


\section{Background and Framework}
\label{background}
In this section we introduce our frame work using a field theory based on Lagrangian mechanics.

\subsection{Deep Neural Networks}
A deep neural network consists of layers of neurons. Suppose that the first layer of neurons with weight matrix $\mathbf{W}_1'$ and bias $\mathbf{b}_1$ takes input $\mathbf{x}_1'$ and outputs $\mathbf{y}_1$
\[
\mathbf{y}_1 = \mathbf{W}_1' \mathbf{x}_1' + \mathbf{b}_1 = \mathbf{W}_1 \mathbf{x}_1,
\]
where $\mathbf{x} = (\mathbf{x}', 1)$ and $\mathbf{W}_1 = (\mathbf{W}'_1, \mathbf{b}_1)$, where $\mathbf{W}_1$ and $\mathbf{b}_1$ are real valued. Now suppose that $R_1$ denotes a nonlinear operator corresponding to a sigmoid or ReLU layer located after the weight layer, so that
\[
\mathbf{y}_1 = R_1 \mathbf{W}_1 \mathbf{x}_1.
\]
For a neural network with $T$ repeating units, the output for layer $t$ is
\begin{eqnarray}
\label{y}
\mathbf{y}_t &=& R_t\mathbf{W}_t \mathbf{y}_{t-1} \nonumber \\
&=& R_t\mathbf{W}_t R_{t-1}\mathbf{W}_{t-1} \ldots R_{1}\mathbf{W}_1 \mathbf{x}_1 \nonumber \\
&=& \bigg( \prod^{t-1}_{n=0} R_{t-n} \mathbf{W}_{t-n}\bigg)\mathbf{x}_1.
\end{eqnarray}

\subsection{Symmetries in Neural Networks}
We now show the necessary and sufficient conditions of preserving symmetry. We explicitly include symmetry transformations in Equation (\ref{y}) and investigate the effects caused by a symmetry transformation of the input in subsequent layers. Suppose $Q_t \in G$ is a transformation matrix in some Lie group $G$ for all $t$. Note that the $Q_t$ are not parameters to be estimated. We write $\mathbf{y}_t = \mathbf{y}_t(Q_t)$, where the dependence on $Q_t$ is obtained from a transformation on the input, $\mathbf{x}_t(Q_t) = Q_t\mathbf{x}_t $, and the weights, $\mathbf{W}_t(Q_t)$
\[
\mathbf{y}_t (Q_t) = R_t \mathbf{W}_t(Q_t)  \mathbf{x}_t(Q_t).
\]
If $G$ is a symmetry group, then $\mathbf{y}_t$ is covariant with $\mathbf{x}_t$, such that $\mathbf{y}_t (Q_t) =Q_t \mathbf{y}_t $. This requires two conditions to be satisfied. First, $\mathbf{W}_t (Q_t) = Q_t \mathbf{W}_t Q_t^{-1}$, where $Q_t^{-1}Q_t = I$ and the existence of the inverse is trivial because $G$ is a group and $Q_t \in G$. The second is the commutativity between $R_t$ and $Q_t$, such that $R_t Q_t = Q_t R_t$. For example, if $g_t \in $ Aff($D$), the group of affine transformations, $R_t$ may not commute with $g_t$. However, commutativity is satisfied when the transformation corresponds to the 2D rotation of feature maps.

Including transformation matrices, the output at layer $t$ is
\[
\mathbf{y}_t(Q_t) = \bigg( \prod^{t-1}_{n=0} R_{t-n} Q_{t-n}\mathbf{W}_{t-n}Q_{t-n}^{-1}\bigg) Q_1 \mathbf{x}_1.
\]

\subsection{The Loss Functional}
Statistical learning requires the loss function to be minimized. It can be written in the form of a mutual information, training error, or the Kullback-Leibler divergence. In this section we approximate the loss function in the continuum limit of samples and layers. Then we define the loss functional to transition into Lagrangian mechanics and field theory. Let  $\mathbf{x}_i = (\mathbf{X}_i,\mathbf{Y}_i) \in \mathcal{X}$ be the $i$-th input sample in data set $\mathcal{X}$, $(\mathbf{X}_i,\mathbf{Y}_i)$ are the features and the desired outputs, respectively, and $i \in \{1, \ldots, N\} $. The loss function is
\[
L = \frac{1}{N} \sum_{i=1}^N L_i( \mathbf{X}_i,\mathbf{Y}_i , \mathbf{W}, \mathbf{Q}),
\]
where $\mathbf{W} = (\mathbf{W}_1,\ldots,\mathbf{W}_T)$,   and  $\mathbf{Q} = (Q_1,\ldots,Q_T)$, $Q_t \in G$ where $G$ is a Lie group, and $T$ is the depth of the network. Taking the continuum limit,
\[
L \simeq \int_{\mathcal{X}} p(\mathbf{X},\mathbf{Y} ) L_{\mathbf{x}}(\mathbf{X},\mathbf{Y}, \mathbf{W}, \mathbf{Q})d\mathbf{X}d\mathbf{Y},
\]
where $ p(\mathbf{X},\mathbf{Y})$ is the joint distribution of $\mathbf{X}$ and $\mathbf{Y}$. Now we define the loss density $ L_{\mathbf{x},t} $ in the continuous layer limit such that 
\[
 L_{\mathbf{x}} \simeq L_{\mathbf{x}}(t=0) + \int_0^{T} \frac{dL_{\mathbf{x}}(\mathbf{X},\mathbf{Y}, \mathbf{W}(t), Q(t))}{dt}dt,
\]
where $L_{\mathbf{x}}(t=0)$ is the value of the loss before training. We let $L_{\mathbf{x},t} = dL_{\mathbf{x}}/dt$ be the loss rate per layer. The loss rate  $ L_{\mathbf{x},t}$ is bounded from below. Therefore 
\[
\min_{\mathbf{W}} L_{\mathbf{x}}( \mathbf{X},\mathbf{Y}, \mathbf{W}, \mathbf{Q}) =  L_{\mathbf{x}}(t=0)+ \int_0^T \min_{\mathbf{W}_t} L_{\mathbf{x},t}(\mathbf{X},\mathbf{Y}, \mathbf{W}(t), Q(t)) dt.
\]
Minimizing the loss rate guarantees the minimization of the total loss. We require $ L_{\mathbf{x},t}$ to be invariant under symmetry transformations. That is, if $Q_1(t) ,Q_2(t) \in G$. Then  
\[
 L_{\mathbf{x},t}( \mathbf{X},\mathbf{Y}, \mathbf{W}(t), Q_1(t)) =  L_{\mathbf{x},t}( \mathbf{X},\mathbf{Y}, \mathbf{W}(t), Q_2(t)).
\]
However if $Q_1(t)$ and $Q_2(t)$ do not belong in the same symmetry group, the above equality does not necessarily hold. Now we define the loss functional for a deep neural network
\[
\mathcal{P}[\mathbf{W},Q] = \int_{\mathcal{X} \times \{0,T\}}p(\mathbf{X},\mathbf{Y})  L_{\mathbf{x},t}( \mathbf{X},\mathbf{Y}, \mathbf{W}(t), Q(t)) d\mathbf{X}d\mathbf{Y} dt.
\]

\subsection{Lagrangian mechanics and Field theory}
Having defined the loss functional now, we can transition into Lagrangian dynamics to give a description of the feature map flow at each layer.  Let the minimizer of the loss rate be 
\[
\mathbf{W}^*(\mathbf{X},\mathbf{Y}, Q(t),t) = \arg \min_{\mathbf{W}} p(\mathbf{X},\mathbf{Y}) L_{\mathbf{x},t}( \mathbf{X},\mathbf{Y}, \mathbf{W}(t), Q(t)).
\]
From now on, we combine $\mathbf{z} = (\mathbf{X},\mathbf{Y})$ as $Y$ only appears in $\mathbf{W}^*$ in this formalism, each $\mathbf{Y}$ determines a trajactory for the feature map representation flow determined by Lagrangian mechanics. Now we define, for each $i$-th element of $\mathbf{W}(t)$, 
\[
v^i(\mathbf{z}, Q(t),t) = W^i(\mathbf{z},Q(t)) - W^{*i}(\mathbf{z}, Q(t),t),
\]
and we set the weights with a non-linear operator $R(t)$ together such that, $w^i(\mathbf{z},Q(t),t) = R(t)v^i(\mathbf{z},Q(t),t)$.
We now define the Lagrangian density, 
\[
\mathcal{L} = \mathcal{L}(\mathbf{z}, \mathbf{w}, \partial_{t}\mathbf{w},\partial_{\mathbf{z}}\mathbf{w},Q(t))
\]
and $\mathcal{L} = \mathcal{T} - \mathcal{V}$, where $\mathcal{T}$ is the kinetic energy and $\mathcal{V}$ is the potential energy. We define the potential energy to be
\[
\mathcal{V} = p(\mathbf{z})  L_{\mathbf{x},t}( \mathbf{z}, \mathbf{W}(t), Q(t)),
\]
The probability density $ p(\mathbf{z}) $ and the loss rate $ L_{\mathbf{x},t}$ are invariant under symmetry transformations. Then $\mathcal{V}$ is an invariant as well. 

\paragraph{Definition: Orthogonal Group} The orthogonal group O($D$) is the group of all $D\times D$ matrices such that $O^T O = I$.

We now set up the conditions to obtain a series expansion of $\mathcal{V}$ around the minimum $w^i(Q_t) = 0$. First, since $\mathcal{V}$ is an invariant. Each term in the series expansion must be an invariant such that $f(w_i(Q_t),w^i(Q_t)) = f(w_i,w^i)$ for all $Q_t \in G$. Suppose $G = O(D)$, the orthogonal group and that $\mathbf{w}^T(Q_t) = \mathbf{w}^T Q^T$ and $\mathbf{w}(Q_t) = Q_t \mathbf{w}$. So $w_iw^i$ is an invariant.  Then $f = w_i w^i$ is invariant for all $Q_t$ where the Einstein summation convention was used  
\[
w_iw^i = \sum_{i=1}^D w_i w^i = \mathbf{w}^T\mathbf{w},
\]
Now we perform a Taylor series expansion about the minimum $w^i = 0$ of the potential,
\[
\mathcal{V} = C + w_i H ^i_j w^j + w_iw_j \Lambda^{ij}_{mn} w^mw^n + O((w_iw^i)^6),
\]
where $H ^i_j = \partial_{w^i} \partial_{w^j} \mathcal{V}$ is the Hessian matrix, and similarly for $\Lambda^{ij}_{mn}$. Because $\mathcal{V}$ is an even function in $w^i$ around the minimum, we must have
\[
 w_i H ^i_j w^j =  \sum_{i=1}^D  w_i H ^i_i w^i.
\]
The O($D$) symmetry enforces that all weight Hessian eigenvalues to be $H^i_i = m^2/2$ for some constant $m^2$. This can be seen in the O(2) case,  with constants $a,b$, $a\neq b$, $Q \in $ O(2) such that $w_1(Q) = w_2$ and $w_2(Q) = w_1$,
\[
a w_1(Q)^2 + b w_2(Q)^2 = a w_2^2 + b w_1^2,
\]
this does not equal $a w_1^2 + b w_2^2$, so the O(2) symmetry implies $a=b$. This can be generalized to the O($D$) case. For the quartic term, the requirement that $\mathcal{V}$ be even around the minimum gives
\[
 w_iw_j \Lambda^{ij}_{mn} w^mw^n = \sum_{i,j=1}^D  \Lambda^{ij}_{ij}w_jw^j w_iw^i.
\]
Similarly the O($D$) symmetry implies $\Lambda^{ii}_{ii} = \lambda/4$ for some constant $\lambda$ and zero for any other elements, the potential is
\[
\mathcal{V} = \frac{m^2}{2} w_iw^i + \frac{\lambda}{4} (w_iw^i)^2,
\]
where the numerical factors were added for convention. The power series is a good approximation in the decoupling limit which may be applicable for Residual Networks.\footnote{ The output of a residual unit is $\mathbf{y}_t = \mathbf{x}_t + \mathbf{F}_t(\mathbf{x}_t, \mathbf{W}_{1t},\mathbf{W}_{2t})$, with $\mathbf{F}_t \ll \mathbf{\mathbf{x}}_t$. Omitting non-linear layers, $\mathbf{F}_t(\mathbf{x}_t, \mathbf{W}_{1t},\mathbf{W}_{2t}) = \mathbf{W}_{t2}\mathbf{W}_{t1}\mathbf{x}_t$. This means the pair of weights in adjacent layers is small also.}
For the kinetic term $\mathcal{T}$, we expand in power series of the derivatives, 
\[
\mathcal{T}= \frac{1}{2}\frac{\partial w_i}{\partial t} \frac{\partial w^i}{\partial t} - \frac{1}{2}\frac{1}{c^2}\frac{\partial w_i}{\partial \mathbf{z}}\frac{\partial w^i}{\partial \mathbf{z}} + O((\partial_t w)^4,(\partial_{\mathbf{z}} w)^4),
\]
 where the coefficient for $(\partial_t w)^2$ is fixed by the Hamiltonian kinetic energy $\frac{1}{2} (\partial_t w)^2$. Higher order terms in $(\partial_t w)^2$ are negligible in the decoupling limit. If the model is robust, then higher order terms in $(\partial_{\mathbf{z}} w)^2$ can be neglected as well. \footnote{The kinetic term $\mathcal{T}$ is not invariant under transformation $Q(t)$. To obtain invariance $\partial_t w^i$ is to be replaced by the covariant derivative $D_t w^i$ so that $(D_t w^i)^2$ is invariant under $Q(t)$, \citep{QFT}. $D_t w^i(\mathbf{z},t,Q(t)) = \partial_t w^i(\mathbf{z},t,Q(t)) + \alpha B(\mathbf{z},t,Q(t))_j^i w^j(\mathbf{z},t,Q(t))$, with $\mathbf{B}(\mathbf{z},t,Q_t) = Q(t)B(\mathbf{z},t)Q(t)^{-1}$. The new fields $\mathbf{B}$ introduced for invariance is not responsible for spontaneous symmetry breaking, the focus of this paper. So we will not consider them further.} The Lagrangian density is \footnote{Formally, the $\partial_{\mathbf{z}}w$ term should be part of the potential $\mathcal{V}$, as $\mathcal{T}$ contains only $\partial_t w$ terms. However we adhere to the field theory literature and put the $\partial_{\mathbf{z}}w$ term in $\mathcal{T}$ with a minus sign, owning to $\mathcal{L} = \mathcal{T} - \mathcal{V}$.}
\[
\mathcal{L} = \frac{1}{2} (\partial_t w)^2 - \frac{1}{2}(\partial_{\mathbf{z}} w)^2 - \frac{m^2}{2} w^2 - \frac{\lambda}{4} (w^2)^2,
\]
where we have set $w^2 = w_i w^i$ and absorbed $c$ into $\mathbf{x}$ without loss of generality. This is precisely the  Lagrangian for a scalar field in field theory. Standard results for a scalar field theory can be found in Appendix \ref{review}. To account for the effect of the learning rate, we employ the results from thermal field theory \citep{thermal} and we identify the temperature with the learning rate $\eta$. So that now $m^2 = -\mu^2 + \frac{1}{4}\lambda \eta^2$, with $\mu^2 > 0$. 

\subsection{Spontaneous Symmetry Breaking}
Consider the following scalar field potential invariant under O($D'$) transformations,
\[
\mathcal{V}(w^i,\eta) =  \frac{m^2(\eta)}{2} w_iw^i + \frac{\lambda}{4} (w_iw^i)^2,
\]
where $m^2(\eta) = -\mu^2 + \frac{1}{4}\lambda \eta^2$, $\mu^2>0$ and learning rate $\eta$. There exists a value of $\eta = \eta_c$ such that $m^2 = 0$. For $\eta > \eta_c$, $w_i^*(\eta > \eta_c) = 0$, where
\[
w_i^*(\eta) = \arg \min_{w^i} \mathcal{V}(w^i,\eta).
\]
The minimum bifurcates at $\eta = \eta_c$ into
\[
w_i^*(\eta < \eta_c) = \pm \sqrt{\frac{-m^2(\eta)}{\lambda}}.
\]
This phenomenon has a profound implication. It is responsible for phase transition in neural networks and generates long range correlation between the input and output. Details from field theory can be found in Appendix \ref{SSB}.

\section{Symmetries in Neural Networks}
\label{symNN}
In this section we show that spontaneous symmetry breaking occurs in neural networks. First, we show that learning by deep neural networks can be considered solely as breaking the symmetries in the weights. Then we show that some non-linear layers can preserve symmetries across the non-linear layers. Then we show that weight pairs in adjacent layers, but not within the same layer, is approximately an invariant under the remnant symmetry leftover by the non-linearities. We assume that the weights are scalar fields invariant under the orthogonal O($D'$) group for some $D'$ and find that experimental results show that deep neural networks undergo spontaneous symmetry breaking.

\paragraph{Theorem: Deep feedforward networks learn by breaking symmetries} {\it Proof}: Let $A_i$ be an operator representing any sequence of layers, and let a network formed by applying $A_i$ repeatedly such that $x_{out} = (\prod^M_{i=1} A_i )x_{in}$. Suppose that $A_i \in$ Aff($D$), the symmetry group of all affine transformations. We have $L =  \prod^D_{i=1} A_i$, where $L \in$ Aff($D$). Then $x_{out} = L x_{in}$ for some $L \in$ Aff($D$) and $x_{out}$ can be computed by a single affine transformation $L$. When $A_i$ contains a non-linearity for some $i$, this symmetry is explicitly broken by the nonlinearity and the layers learn a more generalized representation of the input. $\square$



Now we show that ReLU preserves some continuous symmetries.

\paragraph{Theorem: ReLU reduces the symmetry of an Aff($D$) invariant to some subgroup Aff$(D')$, where $D'< D$.} {\it Proof}: Suppose $R$ denotes the ReLU operator with output $\mathbf{y}_t$ and $Q_t \in$ Aff($D$) acts on the input $\mathbf{x}_t$, where $R(\mathbf{x}) = \max(0,\mathbf{x})$. Let $\mathbf{x}^T \mathbf{x}$ be an invariant under $Q \in$ Aff($D$) and let $\mathbf{x}^T = (\gamma, \nu)$, $\nu <0$ and $\gamma >0$. Let $ \mathbf{a} = R\mathbf{x} = (\gamma,0)$. Then $\mathbf{a}^T \mathbf{a}= \mathbf{x}^T RR \mathbf{x} = \gamma^T \gamma$. Then $\mathbf{a}^T \mathbf{a}$ is an invariant under Aff($D'$) where $D' = \dim \gamma$. Note that $\gamma^i$ can be transformed into a negative value as it has passed the ReLU already.  $\square$

\paragraph{Corollary} If there exist some $Q \in$ G that commutes with a nonlinear operator $R$, such that $QR=RQ$, then $R$ preserves the symmetry $G$.

\paragraph{Definition: Remnant Symmetry} If $Q_t\in G$ commutes with a non-linear operator $R_t$ for all $Q_t$, then $G$ is a remnant symmetry at layer $t$.

For the loss function $ L_i( \mathbf{X}_i,\mathbf{Y}_i , \mathbf{W}, \mathbf{Q})$ to be invariant, we need the predicted output $\mathbf{y}_T$ to be covariant with $\mathbf{x}_i$. Similarly for an invariant loss rate $L_{\mathbf{x},t}$ we require $\mathbf{y}_t$ to be covariant with $\mathbf{x}_t$. The following theorem shows that a pair of weights in adjacent layers  can be considered an invariant for power series expansion.

\paragraph{Theorem: Neural network weights in adjacent layers form an approximate invariant} Suppose a neural network consists of affine layers followed by a continuous non-linearity, $R_t$, and that the weights at layer $t$, $\mathbf{W}_{t}$ transform as $Q_t\mathbf{W}_t Q_t^{-1}$, and that $Q_t \in H$ is a remnant symmetry such that $Q_t R_t = R_t Q_t$. Then 
$\mathbf{W}_t \mathbf{W}_{t-1}$ can be considered as an invariant for the loss rate.

{\it Proof:} 
Consider $\mathbf{x}(Q_t) = Q_t \mathbf{x}_t$, then
\begin{eqnarray}
\mathbf{y}_t(Q_t) & = & R_t \mathbf{W}_t(Q_t) \mathbf{x}_t(Q_t) \nonumber \\
& =& R_t Q_t \mathbf{W}_t Q_t^{-1} Q_t \mathbf{x}_t \nonumber \\
& = & R_t Q_t \mathbf{W}_t \mathbf{x}_t \nonumber \\
& =& Q_t R_t \mathbf{W}_t \mathbf{x}_t, \nonumber 
\end{eqnarray}
where in the last line $Q_t R_t = R_t Q_t$ was used, so $\mathbf{y}_t(Q_t) = Q_t \mathbf{y}_t$ is covariant with $\mathbf{x}_t$. Now, $\mathbf{x}_t = R_{t-1} \mathbf{W}_{t-1} \mathbf{x}_{t-1}$, so that
\[
\mathbf{y}_t(Q_t)  = Q_t R_t \mathbf{W}_t  R_{t-1} \mathbf{W}_{t-1} \mathbf{x}_{t-1}.
\]
The pair $(R_t \mathbf{W}_t)(R_{t-1} \mathbf{W}_{t-1})$ can be considered an invariant under the ramnant symmetry at layer $t$. Recall that $\mathbf{w}_t = R_t (\mathbf{W}_t-\mathbf{W}^*_t)$. Therefore $\mathbf{w}_t \mathbf{w}_{t-1}$ is an invariant. $ \square$

In the continuous layer limit, $\mathbf{w}_t\mathbf{w}_{t-1}$ tends to $\mathbf{w}(t)^T\mathbf{w}(t)$ such that $\mathbf{w}(t)$ is the first layer and $\mathbf{w}(t)^T$ corresponds to the one after. Therefore $\mathbf{w}(t)$ can be considered as a scalar field under the remnant symmetry. The remnant symmetry is not exact in general. For sigmoid functions it is only an approximation. The crucial feature for the remnant symmetry is that it is continuous so that strong correlation between inputs and outputs can be generated from spontaneous symmetry breaking. We will state the Goldstone Theorem from field theory without proof.

\paragraph{Theorem (Goldstone)} For every spontaneous broken continuous symmetry, there exist a weight $\pi$ with zero eigenvalue in the Hessian $m_{\pi}^2=0$. 

In any case, we will adhere to the case where the remnant symmetry is an othogonal group O$(D')$. Note that $\mathbf{W}$ is a $D\times D$ matrix and $D'$. We choose a subset $\mathbf{\Gamma} \in \mathbb{R}^{D'} $ of $\mathbf{W}$ such that $\mathbf{\Gamma}^T\mathbf{\Gamma}$ is invariant under O$(D')$. Now that we have an invariant. We can write down the Lagrangian for a deep feedforward network for the weights responsible for spontaneous symmetry breaking.

\paragraph{The Lagrangian for deep feedforward networks in the decoupling limit} Let $\gamma^i = R( \Gamma^i - \Gamma^{*i})$, using Equation (\ref{lag})

\begin{equation}
\label{lag}
\mathcal{L} = \frac{1}{2} (\partial_t \gamma^i)^2 - \frac{1}{2} (\partial_{\mathbf{z}} \gamma^i)^2 - \frac{m^2(\eta)}{2} \gamma^i \gamma_i - \frac{\lambda}{4}   (\gamma^i \gamma_i)^2.
\end{equation}

Now we can use standard field theory results and apply it to deep neural networks. A review for field theory is given in Appendix \ref{review}. The formalism for spontaneous symmetry breaking is given in Appendix \ref{SSB}.


\section{Main Results}
\label{pred}
Spontaneous symmetry breaking splits the set of weight deviations $\gamma$ into two sets $(\sigma,\pi)$ with different behaviors. Weights $\pi$ with zero eigenvalues and a spectrum dominated by small frequencies $\mathbf{k}$ in its correlation function. The other weights $\sigma$, have Hessian eigenvalues $\mu^2$ as the weights before the symmetry is broken. In Appendix \ref{SSB}, a standard calculation in field theory shows that the correlation functions of the weights have the form
\begin{equation}
\label{corr}
P_{\sigma,\pi} (t, \mathbf{k})  =  \frac{i}{2 \omega_0} \exp\bigg(-i \omega_0 t \bigg),
\end{equation}
where $\omega_0 = \sqrt{\mathbf{|k|}^2+m_{\sigma,\pi}^2}$, $m^2_{\sigma} = -\mu^2_{\sigma} + \frac{1}{4}\lambda \eta^2$ and $m^2_{\pi} =  \frac{1}{4}\lambda \eta^2$. The correlation function of $\pi$ approaches infinity as $\mathbf{k} \rightarrow 0$, as $m_{\pi}^2  \rightarrow 0$. This corresponds to large correlation over all sample space {\it and} layers $t$.

\paragraph{Gradient variance explosion} We can now prove the results found by \citet{opening}. In their Figure 4, they showed that the variance in the weight gradients grow by orders of magnitude near the end of training. We claim that this is due to long range correlations generated by spontaneous symmetry breaking.

Note that during training, the weights or deviation from the mean are given by stochastic gradient descent, 
\[
\gamma^i = - \nabla_w L,
\]
where $L$ is the loss function. 

Suppose that early in the training, spontaneous symmetry breaking has yet to take place. The loss rate is from Equation (\ref{lag})
\[
\mathcal{V} = \frac{m^2}{2} (\gamma^i\gamma_i) + \frac{\lambda}{4}(\gamma^i\gamma_i)^2.
\]
The correlation function \footnote{Please note that this does not have the same definition as the correlation in statistics, please see Appendix \ref{review} for details.} for $\gamma^i$ is localized due to a large Hessian eigenvalue $m^2$. After spontaneous symmetry breaking, the $\pi$ fields with large long range correlation are generated. So at each layer $t$
\[
\frac{\langle \pi\pi \rangle }{\langle \gamma\gamma \rangle } \rightarrow \infty.
\]
This can be seen by fixing $t$ in Equation (\ref{corr}), so $\langle \pi\pi \rangle = P_{\pi}(t,\mathbf{k}) \propto  (|\mathbf{k}| + \frac{1}{4}\lambda \eta^2 )^{-1}$ and $\lambda, \eta \simeq 0$. This means that the gradient variance after spontaneous symmetry breaking blows up as observed by \citet{opening}. Recall that the correlation function corresponds to the weight deviation from the mean across sample space $\mathbf{z}, \mathbf{z'} \in \mathcal{X}$ and layers $t,t'$. A large correlation across sample space means that the feature map representations, which are a product of weights, are independent to the input. Together with a large correlation across the layers, means that the feature maps are highly correlated on the output. This is precisely the behaviors observed by \citet{opening}. 


\paragraph{Robustness of deep neural networks} We find that neural networks are resilient to overfitting. Recall that the fluctuation in the weights can arise from sampling noise. Then $(\partial_{\mathbf{z}} w^i)^2$ can be a measure of model robustness. A small value denotes the weights' resistance to sampling noise. If the network were to overfit, the weights would be very sensitive to sampling error. After spontaneous symmetry breaking, weights with zero eigenvalues obey the Klein-Gordon equation with $m_{\pi}^2 = 0$,
\[
(\partial_{\mathbf{z}} \pi)^2= (\partial_{\mathbf{z}} \pi^*)(\partial_{\mathbf{z}} \pi) = |\mathbf{k}|^2 , \quad \pi = \exp(i\omega t - i \mathbf{k} \cdot \mathbf{z}).
\]
The singularity in the correlation function suggests $|\mathbf{k}|^2 \simeq 0$. The zero eigenvalue weights provide robustness to the model. \citet{understand} referred to this phenomenon as implicit regularization. This is independent of the data distribution. Therefore the network can learn data with random labels, also observed by \citet{understand}.

\paragraph{Weight freeze out} Similarly, we predict that there should be a set of weights that do not change, or very slowly varying, over the layers. This can be seen by $|\partial_t \pi|^2 = \omega^2 $. And $\omega = |\mathbf{k}|$. One could use this to identity the undamped fluctuations. 

\subsection{Validation of Proposed Framework for Neural Networks in the Literature} 
\label{summary}
In this section we summarize experimental findings that can be explained by the proposed field theory and the perspective of symmetry breaking. Here $Q' \in G$ acts on the the input and hidden variables  $\mathbf{x}, \mathbf{h}$, as $Q\mathbf{x}, Q\mathbf{h}$.
\begin{itemize}
\item The shape of the loss function after spontaneous symmetry breaking has the same shape observed by \citet{loss} towards the end of training.
\item Identity mapping outperforms other skip connections \citep{identity} is a result of the residual unit's output being small. Then the residual units can be decoupled leading to a small $\lambda$ and so it is easier for spontaneous symmetry breaking to occur, from $m^2 = -\mu^2 + \frac{1}{4}\lambda \eta^2$.
\item Skip connection across residual units breaks additional symmetry. Suppose now an identity skip connection connects $\mathbf{x}_1$ and the output of $\mathbf{F}_2$. Now perform a  symmetry transformation on $\mathbf{x}_1$ and $\mathbf{x}_2$, $Q_1$ and $Q_2 \in G$ , respectively. Then the output after two residual untis is $Q \mathbf{x_3} = Q_1\mathbf{x_1} + Q_2 \mathbf{x}_2 + Q_2 \mathbf{F}_2$. Neither $Q = Q_1$ nor $Q = Q_2$ can satisfies the covariance under $G$. This is observed by \citet{skip}.
\item The shattered gradient problem \citep{shattered}. It is observed that the gradient in deep (non-residual) networks is very close to white noise. This is reflected in the exponential in Equation (\ref{noise}).  This effect on ResNet is reduced because of the decoupling limit $\lambda \rightarrow 0$. This leads to the weight eigenvalues $m^2$ being larger in non-residual networks owing to $m^2 = -\mu^2 + \lambda \eta^2$. And so a higher oscillation frequency in the correlation function.

\item Spontaneous symmetry breaking generates long range correlations of the weights and feature map representations across all input samples and layers such that feature map representations are almost independent of the input, but highly correlated with the output. This is the result shown in Figure 2 and 3 by \citet{opening}. Further, spontaneous symmetry breaking causes the fluctuation in the weights to grow dramatically and that is reflected in the increase in variance of the gradients during training, shown in Figure 4 of \citet{opening}.
\item In recurrent neural networks, multiplicative gating \citep{gating} combines the input $\mathbf{x}$ and the hidden state $\mathbf{h}$ by an element-wise product. Their method outperforms the method with an addition $\mathbf{x} + \mathbf{h}$ because the multiplication gating breaks the covariance of the output. A transformation $Q\mathbf{x}*Q\mathbf{h} \neq Q(\mathbf{x}*\mathbf{h})$,  whereas for addition  the output remains covariant $Q\mathbf{x}+Q\mathbf{h} = Q(\mathbf{x}+\mathbf{h})$.
\end{itemize}

\section{Concluding Remarks}
In this work we solved one of the most puzzling mysteries of deep learning by showing that deep neural networks undergo spontaneous symmetry breaking. This is a first attempt to describe a neural network with a scalar quantum field theory. We have shed light on many unexplained phenomenon observed in experiments. From the observation that deep networks can learn random training labels with no error \citep{understand}, well generalized models have many zero valued weights \citep{eigen}, to the phenomenon that feature maps becomes highly correlated with the output in early layers \citep{opening}. These are all direct consequences of spontaneous symmetry breaking and their experimental results confirm our theory. There are also other experiments that validates our theory. The observation that gradients are highly oscillatory across the layers in a network without skip connections \citep{shattered} and that the oscillation is reduced in ResNets confirm the form of the correlation function in Equation (\ref{corr}).

We believe that this work provides an understanding of deep neural network which will motivate further experimental and theoretical investigations. We have shown that we can predict the performance of a deep network by considering the symmetry in the layers. This gives practitioners a guide to design better performing architectures more efficiently. We have also provided an alternate framework to information theory which can be experimentally tested against.

\appendix

\section{Review of Field Theory}
\label{review}
In this section we state the relevant results in field theory without proof. We use Lagrangian mechanics for fields $w(\mathbf{x},t)$. Equations of motion for fields are the solution to the Euler-Lagrange equation, which is a result from the principle of least action. The action, $S$, is
\[
S[w] = \int L(t,w, \partial_t w) dt,
\]
where $L$ is the Lagrangian. Define the Lagrangian density
\[
L(t) = \int \mathcal{L}(\mathbf{z},t , w, \partial_{\mathbf{z},t}w) d\mathbf{x}.  
\]
The action in term of the Lagrangian density is
\[
S[w] = \int \mathcal{L}(\mathbf{z},t , w, \partial_{\mathbf{z},t} w) d\mathbf{x} dt.
\]
The Lagrangian can be written as a kinetic term $\mathcal{T}$, and a potential term $\mathcal{V}$ (loss function),
\[
\mathcal{L} = \mathcal{T} - \mathcal{V}
\]
For a scalar field $w(\mathbf{x},t)$,
\[
\mathcal{T} = \frac{1}{2}\bigg(\frac{\partial w}{\partial t}\bigg)^2 - \frac{1}{2}\frac{1}{c^2}\bigg(\frac{\partial w}{\partial \mathbf{z}}\bigg)^2 = (\partial_t w)^2  - (\partial_{\mathbf{z}} w)^2
\]
where we have set the constant $c^2 = 1$ without loss of generality. The potential for a scalar field that allows spontaneous symmetry breaking has the form
\[
\mathcal{V} = \frac{m^2}{2} w^2 + \frac{\lambda}{4} w^4.
\]
In the decoupling limit, $\lambda \rightarrow 0$, the equation of motion for $w$ is the Klein-Gordon Equation
\[
 [(\partial_t )^2 - (\partial_{\mathbf{z}} )^2 - m^2] w = 0.
\]

In the limit of $m^2 \rightarrow 0$, the Klein-Gordon Equation reduces to the wave equation with solution
\[
w(\mathbf{z},t) = e^{i(\omega t - \mathbf{k} \cdot \mathbf{z})},
\]
where $i = \sqrt{-1}$.

One can treat $w$ as a random variable. The probability distribution of the scalar field $w(\mathbf{x},t)$ is $p[w] = \exp(-S[w])/\mathcal{Z}$, where $\mathcal{Z}$ is some normalizing factor. The distribution peaks at the solution of the Klein-Gordon equation since it minimizes the action $S$. Now we can define the correlation function between $w(\mathbf{z}_1, t_1)$ and $w(\mathbf{z}_2, t_2)$, 
\[
\langle w(\mathbf{z}_1, t_1) w(\mathbf{z}_2, t_2) \rangle = \frac{1}{\mathcal{Z}}\int  w(\mathbf{z}_1, t_1) w(\mathbf{z}_2, t_2) e^{-S[w]} \mathcal{D}w,
\]
where $\int \mathcal{D} w$ denotes the integral over all paths from $(\mathbf{z}_1, t_1) $ to $(\mathbf{z}_2, t_2)$. In the decoupling limit $\lambda \rightarrow 0$, it can be shown that
\[
\exp(-S[w]) = \exp\bigg(-\int  w (\partial_t^2 - \partial_{\mathbf{z}}^2 + m^2) w \  d\mathbf{x} dt\bigg),
\]
where Stokes theorem was used and the term on the boundary of (sample) space is set to zero. The above integral in the exponent is quadratic in $w$ and the integral over $\mathcal{D}w$ can be done in a similar manner to  Gaussian integrals. The correlation function of the fields across two points in space and time is
\[
\langle w(\mathbf{z}_1, t_1) w(\mathbf{z}_2, t_2) \rangle = G(\mathbf{z}_1, t_1,\mathbf{z}_2, t_2),
\]
where $G(\mathbf{z}_1, t_1,\mathbf{z}_2, t_2)$ is the Green's function to the Klein-Gordon equation, satisfying
\[
 (\partial_t^2 - \partial_{\mathbf{z}}^2 + m^2) G(\mathbf{z}_1, t_1,\mathbf{z}_2, t_2) = \delta(\mathbf{z}_1 - \mathbf{z}_2) \delta(t_1 - t_2).
\]
The Fourier transformation of the correlation function is
\[
G(\omega, \mathbf{k}) = \frac{1}{\omega^2 - |\mathbf{k}|^2 - m^2}, \quad m^2 > 0.
\]
An inverse transform over $\omega$ gives
\[
G(t, \mathbf{k}) = \frac{i}{2\omega_0}e^{-i\omega_0 t}, 
\]
with $\omega_0^2 = |\mathbf{k}|^2 + m^2$.

\section{Spontaneous Symmetry Breaking in the Orthogonal Group O($D'$)}
\label{SSB}
In this section we show that weights $\pi$ with small, near zero, eigenvalues $m_{\pi}^2 = \frac{1}{4}\lambda \eta^2$ are generated by spontaneous symmetry breaking. Note that we can write the Lagrangian in Equation (\ref{lag}) as $\mathcal{L} = \mathcal{T} - \mathcal{V}$. Consider weights $\gamma$ that transforms under O$(D')$, from Equation (\ref{lag})
\begin{eqnarray}
\mathcal{T} &=&  \frac{1}{2}(\partial_t \gamma^i)^2 - \frac{1}{2}(\partial_{\mathbf{z}} \gamma^i)^2, \nonumber \\
\mathcal{V} &=& \frac{m^2}{2} \gamma^i \gamma_i + \frac{\lambda}{4}   (\gamma^i\gamma_i)^2.
\end{eqnarray}

When $m^2 = -\mu^2 + \frac{1}{4}\lambda \eta^2<0$, it can be shown that in this case the loss minimum is no longer at $\gamma^i = 0$, but it has a degenerate minima on the surface such that $ \sum_i (\gamma_i)^2 = v$, where $v = \sqrt{-m^2/\lambda}$. Now we pick a point on this loss minima and expand around it. Write $\gamma^i = (\pi^k, v + \sigma)$, where $k \in \{1,\ldots, D'-1\}$. Intuitively, the $\pi^k$ fields are in the subspace of degenerate minima and $\sigma$ is the field orthogonal to $\pi$. Then it can be shown that the Lagrangian can be written as
\[
\mathcal{L} = \mathcal{T}_{\pi} + \mathcal{T}_{\sigma} - \mathcal{V}_{\pi} - \mathcal{V}_{\sigma} - \mathcal{V}_{\pi \sigma},
\]
where, in the weak coupling limit $\lambda \rightarrow 0$,
\begin{eqnarray}
\mathcal{T}_{\pi} &=& \frac{1}{2}(\partial_t \pi^k)^2 - \frac{1}{2}(\partial_{\mathbf{x}} \pi^k)^2, \nonumber \\ 
\mathcal{T}_{\sigma} &=& \frac{1}{2}(\partial_t \sigma)^2 - \frac{1}{2}(\partial_{\mathbf{x}} \sigma)^2, \nonumber \\ 
\mathcal{V}_{\pi} & = & O(\lambda), \nonumber \\
\mathcal{V}_{\sigma} & = & - m^2 \sigma^2,\nonumber \\
\mathcal{V}_{\pi \sigma} & = & O(\lambda), \nonumber \\
\end{eqnarray}
the fields $\pi$ and $\sigma$ decouple from each other and can be treated separately. The $\sigma$ fields satisfy the Klein-Gordon Equation $( \square- m^2)\sigma = 0$, with $\square = \partial_t^2 - \partial_{\mathbf{z}}^2$. The $\pi$ fields satisfy the wave-equation, $\square \pi = 0$. The correlation functions of the weights across sample space and layers, $P_{\sigma} =\langle \sigma(\mathbf{z}',t') \sigma(\mathbf{z},t) \rangle$ and $P_{\pi} = \langle \pi(\mathbf{z}',t') \pi(\mathbf{z},t) \rangle$ are the Green's functions of the respective equations of motion. Fourier transforming the correlation functions give
\begin{equation}
\label{noise}
P_{\sigma,\pi} (t, \mathbf{k})  =  \frac{i}{2 \omega_0} \exp\bigg(-i \omega_0 t \bigg),
\end{equation}
where $\omega_0 = \sqrt{|\mathbf{k}|^2+|m_{\sigma, \pi}^2|}$, and $m_{\pi}^2=\frac{1}{4}\lambda \eta^2 \simeq 0$. The correlation function $P_{\pi}$ is dominated by values of $|\mathbf{k}| \simeq 0$. Therefore $\langle \pi\pi \rangle \rightarrow \infty$ as $\lambda\eta^2 \rightarrow 0$. On the other hand, it can be shown that $\langle \sigma\sigma \rangle$ is damped by the weight eigenvalues $|m^2|$. The singularity in the correlation function means that the value of the weights at the start of the layer is highly correlated with the ones in later layers.

In the language of group theory. The O($D$) symmetry is broken down to O$(D-1)$. Elements of O($D$) are the $D\times D$ orthogonal matrices, which have $D(D-1)/2$ independent continous symmetries (e.g. the Euler angles in $D=3$). The number of continuous broken from O($D$) to O($D-1$) is $D-1$. In the above example we showed that this corresponds to the $D-1$ $\pi^k$ fields. Each of which have infinite correlation functions.

Even though we formulated our field theory based on the decoupling limit of ResNets, the result of infinite correlation is very general and can be applied even if the decoupling limit is not valid. It is a direct result of spontaneous symmetry breaking. We state the Goldstone Theorem without proof.

\paragraph{Theorem (Goldstone):} For every continuous symmetry that is spontaneously broken, a weight $\pi$ with corresponding eigenvalue $m_{\pi}^2 = 0$ is generated at zero temperature (learning rate $\eta$). $\square$

\bibliography{iclr2018_conference}
\bibliographystyle{iclr2018_conference}

\end{document}